\documentclass[runningheads]{llncs}
\usepackage{graphicx}
\usepackage[usenames, dvipsnames]{xcolor}

\begin{document}
\title{Evaluation of the importance of criteria for the selection of cryptocurrencies}
%
%
\author{
Natalia A. Van Heerden\inst{1} \and
Juan B. Cabral\inst{2,3} \and
Nadia A. Luczywo\inst{4,5,6}
}
\authorrunning{Van Heerden et al.}
%
\institute{
    Universidad Blas Pascal, C\'ordoba, Argentina 
    \and
    Centro Internacional Franco Argentino de Ciencias de la Informaci\'on y de Sistemas, CONICET--UNR, Argentina 
    \and
    Instituto de Astronom\'ia Te\'orica y Experimental, CONICET--UNC, Argentina
    \and
    Facultad de Ciencias Econ\'omicas, UNC, Argentina
    \and
    Laboratorio de Ingenier{\'\i}a y Mantenimiento Industrial, Facultad de Ciencias Exactas, F\'{\i}sicas y Naturales, UNC, 
    Argentina
    \and
    Facultad de Ciencias de la Administración, UNDE, Argentina
\email{nativanheer@gmail.com, \{jbcabral, nluczywo\}@unc.edu.ar}}

\maketitle              
\begin{abstract}
In recent years, cryptocurrencies have gone from an obscure niche to a prominent place, with investment in these assets becoming increasingly popular. However, cryptocurrencies carry a high risk due to their high volatility. In this paper, criteria based on historical cryptocurrency data are defined in order to characterize returns and risks in different ways, in short time windows (7 and 15 days); then, the importance of criteria is analyzed by various methods and their impact is evaluated. Finally, the future plan is projected to use the knowledge obtained for the selection of investment portfolios by applying multi-criteria methods.

\end{abstract}
\section{Introduction}
\label{sect:intro}

Cryptocurrencies, also known as ``electronic currencies'' or ``virtual currencies'', are becoming a popular alternative  of payment and exchange \cite{hileman2017global}. Investors choose these digital assets not only as a value storage, but also to diversify their assets portfolio, relying on the digital and cryptographic aspect to secure and manage transactions with the high --as well as volatile-- returns of these assets \cite{walther2019exogenous}.

Since the emergence of Bitcoin \cite{nakamoto2008bitcoin}, when the lack of knowledge of blockchain technology \cite{chaum1979computer} was at its peak, thousands of new virtual currencies have emerged. Due to the increasing number number of available cryptocurrencies \cite{elbahrawy2017evolutionary}, and in order to take advantage of the returns derived from these assets, all while considering the potential risks, it is interesting to structure a decision support process aimed at the selection of cryptocurrencies in order to invest in them.

The inefficiencies in the information available on cryptocurrencies characterize the market as volatile and uncertain, where there are considerable risks \cite{bariviera2017inefficiency}. The precautions that an investor takes for proper asset management play a fundamental role. The need arises then, to optimize resources and to make coherent decisions in conditions of conflict and uncertainty, which incorporate decision analysis techniques.

Among the quantitative tools available for decision-making are the discrete multi-criteria decision support methods (MCDA), which allow working with multiple evaluation criteria simultaneously, and identifying the relative importance of each one in order to evaluate between different alternatives \cite{alberto2013apoyo}. Therefore, this work proposes to structure the cryptocurrency investment decision problem for short windows of 7 and 15 days, extracting the criteria that characterize returns and risks from historical data based on time series of quotes, and evaluating the importance of the criteria. obtained and their impact on MCDA models.

This paper is organized as follows: Section~\ref{sect:methodology} provides a brief description of the criteria-weighting and multi-criteria methodologies used. Section~\ref{sect:applitaction} determines the design of the criteria, the selection of alternatives and the application of methodologies detailed in Section~\ref{sect:methodology}. Finally, Section~\ref{sect:conclusions} presents the conclusions and future perspectives.

\section{Methodology}
\label{sect:methodology}

In order to assign importance to the criteria of the experiment, four different methodologies were used. Likewise, although the main objective of this work consists of evaluating the importance of the criteria obtained, a performance qualification was performed using a multi-criteria control method; Due to its simplicity and interpretability, TOPSIS \cite{hwang1981methods} was chosen. 

\subsection{Importance of the criteria}

For the weighting of the importance of the criteria, the strategies detailed below were chosen.

\begin{enumerate}
    \item \textbf{Mean weights:} a normalized constant as weight was assigned to all criteria, using the formula:
    
    $$ w_j = \frac{1}{m} $$
        
    where $w_j$ is the weight of criterion $j$ and $m$ is the number of criteria under analysis.
    \item \textbf{Standard deviation}: is a method based only on the intensity of the contrast of the criteria, by means of the expression:
    
    $$ w_j = \frac{\sigma_j}{\sum_{k=1}^{m}\sigma_k} $$
    
    where $w_j$ is the weight of the criterion $j$, $\sigma_j$ is the standard deviation of the scores of the criterion $j$, $m$ is the number of criteria and $\sigma_k$ is the standard deviation of scores of the $k$ criterion.
    \item \textbf{Entropy:}  is considered as a measure of the information contained in an information source. The following formula is used for its calculation:
    
    $$ H_j = \sum_{i=1}^{n} p_i \log{\frac{1}{p_i}} $$
    
    where $H_j$ is the entropy of the criterion $j$, $p_i$ is the probability that the this criteria value will appear, and $n$ is the total number of criteria. After establishing the entropy values, they are normalized under the following expression:
    
    $$ w_j = \frac{H_j}{\sum_{k=1}^{m}H_k} $$
    
    where $w_j$ is the weight of the criterion $j$, $m$ is the number of criteria and $H_k$ is the weight of the $k$-th criterion.
    \item \textbf{CRITIC:} This method, an acronym for \textit{CRiteria Importance Through Intercriteria Correlation} \cite{diakoulaki1995determining}, weights each criterion according to the expression:
    
    $$ C_j = \sigma_j \sum_{k=1}^{m} (1 - r_{jk}) $$ 
    
    where $C_j$ is the weight of the $j$-th criterion, $\sigma_j$ is the standard deviation of criterion $j$, $m$ is the number of criteria and $r_{jk}$ is the Pearson correlation coefficient between criteria $j$ and $k$. The higher the value of $C_j$, the greater the amount of information transmitted by the corresponding criterion and the greater its relative importance for the decision-making process. After establishing the values, they are normalized under the following expression:
    
    $$ w_j = \frac{C_j}{\sum_{k=1}^{m}C_k} $$
    
    where $w_j$ is the weight of criterion $j$ and $C_k$ is the weight or weight of criterion $k$.
    
\subsection{TOPSIS}

The ``\textit{Technique for Order of Preference by Similarity to Ideal Solution}'' (TOPSIS) is a multi-criteria decision analysis method developed by Hwang and Yoon in 1981 \cite{hwang1981methods}. The method is based on the concept of similarity between the ideal alternative and an anti-ideal one, since it considers desirable that an alternative be located at the shortest distance from the ideal solution and the greatest distance from the anti-ideal solution.

In the TOPSIS method, a similarity ratio is defined, which assesses the performance of each alternative such that if the alternative is closer to the ideal point it will be closer to 1, and on the contrary, if it is closer to the anti-ideal point, its value will be closer to 0. The following ratio is applied to its calculation:

$$ C_i = \frac{S_i^{-}}{S_i^{+} + S_i^{-}} $$

where $C_i$ is the similarity index of alternative $i$, $S_i^+$ the Euclidean distance of $i$ with respect to the ideal value and $S_i^-$ the Euclidean distance of $i$ with respect to the anti-ideal value.
\end{enumerate}

\section{Application}
\label{sect:application}

The experiment consisted of the extraction of criteria from historical time series of nine cryptocurrencies: Cardano, Binance Coin, Bitcoin, Dogecoin, Ethereum, Chainlink, Litecoin, Stellar and XRP.

\subsection{Data treatment and criteria definition}

The data analyzed comes from the historical data set ``Cryptocurrency Historical Prices''\footnote{\url{https://www.kaggle.com/sudalairajkumar/cryptocurrencypricehistory}} retrieved on July 21st, 2021. Two decision matrices were created for two sizes of overlapping moving windows: 7 and 15 days. Six criteria were defined on these windows that seek to represent returns and risks and investment:

\begin{enumerate}
    \item average Window return ($\bar{x}RV$) - Maximize: is the average of the differences between the closing price of the cryptocurrency on the last day and the first day of each window, divided by the price on the first day;
    \item window return deviation ($sRV$) - Minimize: is the standard deviation of window return. The greater the deviation, the returns within the windows have higher variance and are unstable.
    \item average of the volume of the window ($\bar{x}VV$) - Maximize: it is the average of the summations of the transaction amount of the cryptocurrency in dollars in each window, representing a liquidity measure of the asset;
    \item window volume deviation ($sVV$) - Minimize: it is the deviation of the window volumes. The greater the deviation, the volumes within the windows have higher variance and are unstable.
    \item mean of the slope ($\bar{x}m$) - Maximize: it is the mean of the slope of the linear trend between the closing prices in dollars and the volumes traded in dollars of the cryptocurrency within each window.
    \item mean of the correlation coefficient ($\bar{x}R^2$) - Maximize: it is the mean of the $R^2$ of the fit of the linear trends with respect to the data. It is a measure that defines how well it explains that linear trend to the data within the window.
\end{enumerate}

\subsection{Alternatives selection}

The selection of the alternatives was based on the ranking of the 20 cryptocurrencies with the highest market capitalization calculated on the basis of the circulating supply, according to the information retrieved from ``All Cryptocurrencies''\footnote{\url{https://coinmarketcap.com/all/views/all/}} on August 4th, 2021. From this list the cryptocurrencies for which there was data for the period from October 9th, 2018 to July 6th, 2021 were taken, and the so-called stablecoins are excluded, since they maintain a stable price and therefore do not carry associated returns. The selected alternatives were then: ADA - Cardano, BNB - Binance Coin, BTC - Bitcoin, DOGE - Dogecoin, ETH - Ethereum, LINK - Chainlink, LTC - Litecoin, XLM - Stellar and XRP - XRP.

\subsection{Determination of the relative weights of the criteria}

\begin{figure}
\includegraphics[width=\textwidth]{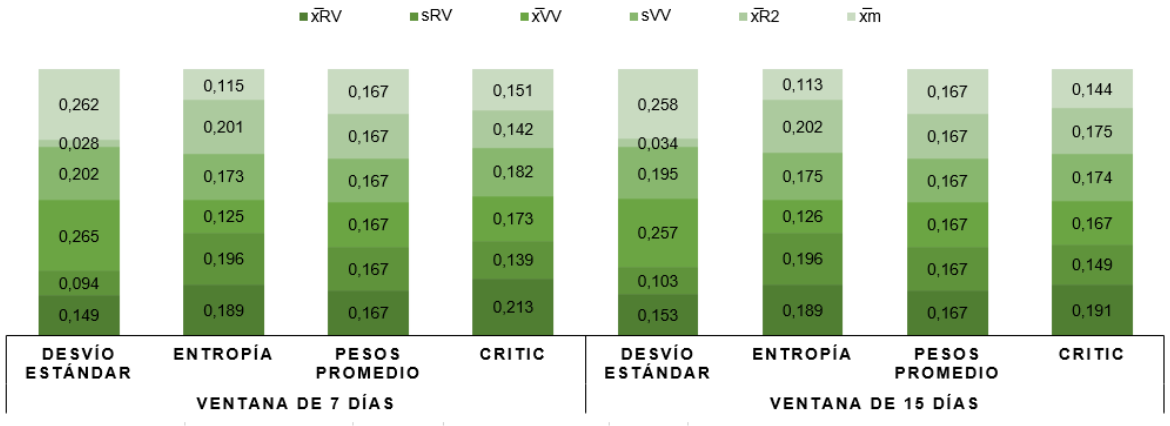}
\caption{Distribution of criteria weights according to calculation methodology.} \label{fig:weights}
\end{figure}

To determine the relative weight, importance or influence of each of the criteria on the final result, four procedures were used: standard deviation, information entropy, average weight and CRITIC. Previously, the evaluations of the alternatives corresponding to minimization criteria were transformed, and then the data were normalized using the vector modulus procedure to be able to calculate the relative weights; The results of the normalization of the alternatives and weights are shown in Table~\ref{table:data}. Likewise, Figure~\ref{fig:weights} shows the distribution of the weights obtained for each of the criteria according to the methodologies used.

\begin{table}[]
    \centering
    \includegraphics[width=\textwidth]{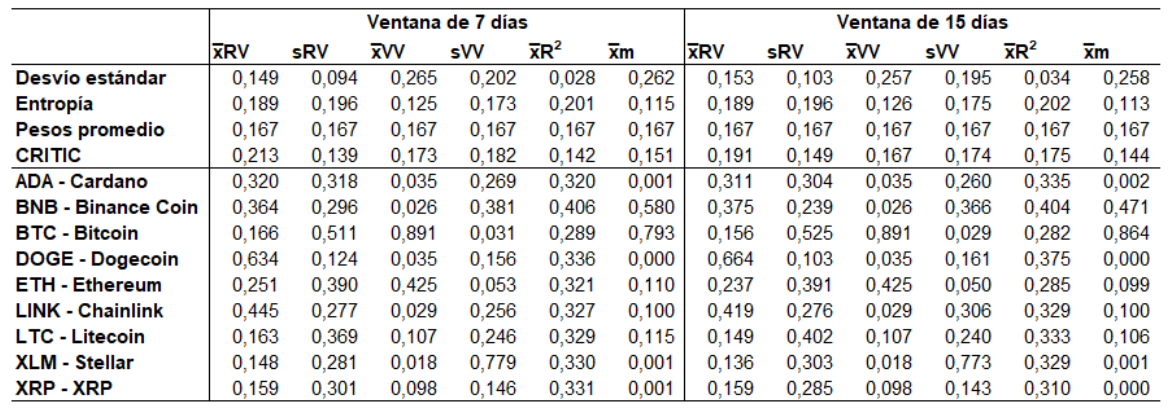}
    \caption{Alternatives and criteria weights according to calculation methodology, normalized.}
    \label{table:data}
\end{table}

\subsection{Impact evaluation of weight selection methods}

As part of the final criteria analysis, we decided to evaluate the impact of the different weight selection methods using TOPSIS because it is easy to interpret through its similarity index, which provides more detailed information than the ranking.

\begin{figure}
\includegraphics[width=\textwidth]{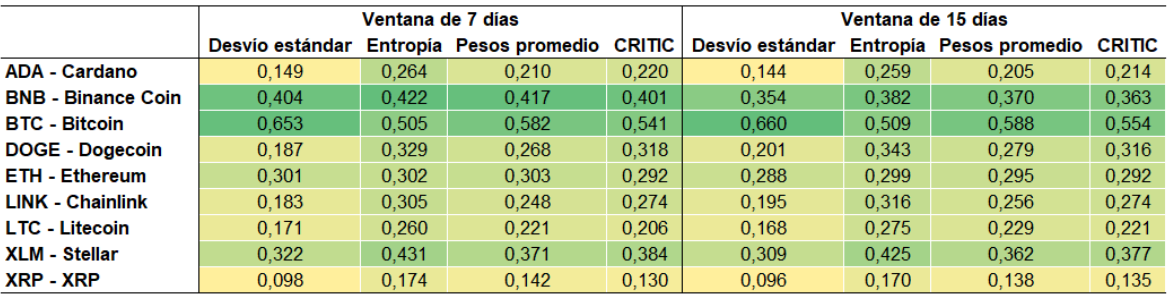}
\caption{TOPSIS similarity indices according to calculation methodology.} \label{fig:results}
\end{figure}

It can be seen in Figure~\ref{fig:results} that the weights obtained through the average weight, entropy and CRITIC methods generate little variation in the order in which an alternative is chosen. Another important result is the little difference between the values obtained between the two window sizes. Already as a collateral result it can be seen that BTC and BNB detach in all cases from the other alternatives.
Finally, we have made public all the results and calculations in a database created to complement this work \cite{dataset}.

\section{Conclusions}
\label{sect:conclusions}

In this work, six criteria were designed from the historical time series of nine cryptocurrencies, and we analyzed them to determine their relative importance in short investment windows. It was determined that under the given conditions and with the existing limitations regarding the amplitude in the time series windows, the weights / importances obtained through the average weight, entropy and CRITIC methods are similar.

As future work, it is proposed to incorporate criteria that take advantage of the information on the market capitalization of cryptocurrencies (available in the time series), use other methods to determine more relative weights of the criteria, perform a sensitivity analysis of the weights and / or evaluate them with different multi-criteria methods beyond TOPSIS.

\section*{Acknowledgments}
The authors would like to thank Martín Beroiz for the help with corrections during the English translation.

%
%
\bibliographystyle{splncs04}
\bibliography{main.bib}

\end{document}